\documentclass[fleqn,usenatbib]{mnras}

\usepackage{newtxtext,newtxmath}
\usepackage{gensymb}
\usepackage{multirow}

\usepackage[T1]{fontenc}
\usepackage{ae,aecompl}

\usepackage{graphicx}	
\usepackage{amsmath}	
\usepackage{amssymb}	
\usepackage{soul}

\title[eROSITA Detection Rates for WTDEs]{eROSITA Detection Rates for Tidal Disruptions of White Dwarfs by Intermediate Mass Black Holes}

\author[A. Malyali et al.]{
A. Malyali,\thanks{E-mail: amalyali@mpe.mpg.de}
A. Rau,
K. Nandra
\\
Max-Planck-Institut f\"ur extraterrestrische Physik,  Giessenbachstrasse 1, 85748 Garching, Germany
}

\date{Accepted XXX. Received YYY; in original form ZZZ}

\pubyear{2019}

\begin{document}
\label{firstpage}
\pagerange{\pageref{firstpage}--\pageref{lastpage}}
\maketitle

\begin{abstract}
White dwarf-black hole tidal disruption events (herein WTDEs) present an opportunity to probe the quiescent intermediate mass black hole population in the universe. We run an extensive set of Monte-Carlo based simulations to explore SRG/eROSITA's detection sensitivity to WTDEs as a function of black hole mass, redshift and time offset between event flaring and it first being observed. A novel estimate of WTDE rate densities from globular clusters and dwarf galaxies is also presented. We combine this with estimated detection sensitivities to infer the rate of eROSITA detecting these events. Depending on the estimate of the intrinsic rate of WTDEs, we anticipate that eROSITA may detect $\sim 3$ events over its 4 year all-sky survey. eROSITA will be most sensitive to systems with black hole masses above $10^4M_{\odot}$, and is most likely to catch these within 5 days of flaring. 
\end{abstract}




\section{Introduction}
Intermediate-mass black holes (IMBHs), with masses in the range $10^3-10^5M_\odot$, are predicted to reside at the centres of globular clusters \citep{Colbert11999,Fabbiano2001,Gultekin2004} and dwarf galaxies \citep{Dong2007,Greene2012}, yet current observational evidence for the existence of IMBHs is uncertain (see \citealt{Mezcua2017} for a review).

One potential avenue for probing the quiescent IMBH population is through Tidal Disruption Events (TDEs), whereby perturbations to a star's orbit lead to it passing too close to the BH and being ripped apart by tidal forces \citep{Hills1975,Lacy1982,Rees1988}. Stellar populations surrounding IMBHs are currently weakly constrained observationally, although it is expected that the rate of tidal disruption is higher for main sequence (MS) than white dwarf (WD) stars \citep{Ramirez-Ruiz2009}. However, the shorter timescales for disruption of WDs relative to MS stars, coupled with potentially higher peak luminosities, is likely to lead to tidal disruptions of WDs being more easily observed and taking up a larger fraction of observed events for IMBHs \citep{MacLeod2014}.

WDs can only be disrupted by black holes with mass $\lesssim 10^5$ $M_{\odot}$ and are swallowed whole by black holes more massive than this, with gravitational radiation likely being the only observational signature in such cases \citep{Luminet1989,East2014}. After disruption, accretion of the bound stellar debris onto the central BH leads to a soft X-ray flare of thermal radiation -- for a review of TDE observations see \citet{Komossa2015} and references therein. Theoretical and computational work has predicted a wealth of additional signatures such as type 1a supernovae-like optical transients due to thermonuclear burning triggered by extreme tidal compression of the white dwarf perpendicular to the orbital plane \citep{LuminetJ1989,Rosswog2008,Rosswog2009,Haas2012,Kawana2017}, modulation of the luminosity output due to changes in accretion rate produced by elliptical orbit behaviour of the WD \citep{Zalamea2010}, and the launching of relativistic jets \citep{DeColle2012, Krolik2012, Shcherbakov2013}. A number of candidates for WTDEs have already been reported; some with X-ray and radio jet signatures \citep{Krolik2011,Shcherbakov2013} and several with non-jetted X-ray signatures \citep{Jonker2013,Glennie2015}. The rapid decay rates involved (the whole flaring episode occurs over hours to weeks depending on WTDE configuration, as opposed to months to years for MS TDEs -- see Fig. 11 of \citealt{Law-Smith2017} for  comparison of different TDE timescales) makes confident classification of WTDEs difficult as there is usually insufficient multi-wavelength evidence to choose one model over competing models, such as flaring stars or off-axis GRBs.

eROSITA (extended Roentgen Survey with an Imaging Telescope Array) \citep{Merloni2012,Predehl2016}, which is on board the Russian-German Spektrum-Roentgen-Gamma (SRG) mission, was launched in July 2019. Its first four years of operation will be dedicated to an X-ray all-sky survey where it will be $\sim$20 times more sensitive than its predecessor, ROSAT, in the soft X-ray band ($0.5-2$ keV) and will be the first imaging survey to cover the whole hard X-ray (2-10keV) sky.
\citet{Khabibullin2014} explored the rate of eROSITA detecting tidal disruption flares of main sequence (MS) stars by supermassive black holes and estimated $\sim 10^3$ new TDE candidates per all sky scan (ie. every six months) -- see also \citet{Thorp2018}. However, the more explosive, shorter timescales of WTDEs has yet to be explored and given the extent of theoretical work put into predicting the observational signatures of these transients, it is useful to understand eROSITA's detection sensitivity to these events.

In this work, we explore eROSITA's detection sensitivity to WTDEs through an extensive set of Monte-Carlo based simulations. Based on previous theoretical and computational work we explore the observational signatures of these events, focusing on non-jetted signatures since the underlying jet formation mechanisms are not well understood. We introduce the simulation framework and underlying theory that we use to model WTDEs and associated X-ray observational properties in Section~\ref{sec:simulations}. eROSITA's detection sensitivity to these events for different WD-BH configurations is explored via simulations in Section~\ref{sec:detection_sensitivity}. This is then combined with the intrinsic rate of WTDEs in the local universe to estimate eROSITA's detection rate in Section~\ref{sec:rate_estimate}. We discuss caveats of our modeling and draw up an approach for multiwavelength follow-up in Section~\ref{sec:discussion}, before presenting conclusions in Section~\ref{sec:conclusions}.

\section{Simulating \lowercase{e}ROSITA observations}
\label{sec:simulations}

Following launch, the SRG satellite will enter an orbit around the Second Lagrangian (L2) point of the Earth-Sun system, where eROSITA will perform eight All-Sky Surveys (eRASS 1-8) during its first four years. The satellite will rotate once every four hours around an axis pointed a few degrees away from the Sun, with observations moving by 1 degree per day (at the equator) perpendicular to this axis. Point sources will pass through eROSITA's $1\degree$ diameter FOV six times per day. With each passage lasting $\sim$40s, this will lead to $\sim$240s exposure for each source per eRASS. We note that
this is a minimum exposure for a sky region. Due to the non-uniformity of exposure across the sky, the Ecliptic poles will be visited more frequently and sources within these regions will have longer exposures \citep{Merloni2012}. This scanning strategy allows eROSITA to probe variability on timescales from $\sim40$s, to days and up to years, albeit with an inhomogeneous and often sparse light curve coverage.

We used the Monte Carlo based code SIXTE\footnote{\url{http://www.sternwarte.uni-erlangen.de/research/sixte/}} \citep{Schmid2012} to simulate eROSITA observations of patches of the sky during the first all-sky survey, eRASS1, where we utilise a spacecraft attitude file to model the time-dependence of eROSITA's pointing direction\footnote{Whilst the final strategy could change for operational reasons, it is expected to be very similar.}. SIXTE requires specification of an instrument description file, and an instrument-independent sky model contained in a \texttt{simput} file\footnote{\url{https://www.sternwarte.uni-erlangen.de/research/sixte/simput.php}}, with details of each source's sky position, flux in a reference energy band, X-ray spectrum and light curve (LC). A photon population is then simulated given this sky model and propagated through the instrument model to produce a set of simulated event files. Whilst computationally demanding, this approach has the advantage that it realistically models instrument specific effects on photon propagation and detection which cannot be modelled analytically.

The sky models used in this work consist of a population of WTDEs, an AGN population, and a soft X-ray background (SXB) originating from collisionally-ionized diffuse gas within the Solar system and Galaxy. Details of each of these components are provided in Sections~\ref{sec:tdes},~\ref{sec:agn_population} and ~\ref{sec:xrb} respectively. We also add a particle background component that is uniform over the detectors (not passing through eROSITA's mirror system) and implemented within the SIXTE simulator based on \citet{Tenzer2010Geant4Background}. For simplicity, we do not include galaxy cluster and stellar populations, but discuss the effects of this omission in Section~\ref{sec:discussion}.

\subsection{WTDEs}
\label{sec:tdes}
\subsubsection{Temporal properties}
A star will be tidally disrupted if it passes too close to a BH. For a system containing a BH of mass $M_{\text{bh}}$ and a star with mass and radius of $M_{\text{*}}$ and $R_{\text{*}}$, respectively, this occurs for stellar orbits where the pericenter radius, $r_{\text{p}}$, is less than the tidal radius \citep{Hills1975}, approximately defined as:
\begin{equation} \label{eqn:r_tidal}
r_{\text{t}}\simeq \left( \frac{M_{\text{bh}}}{M_{\text{*}}} \right)^{1/3}R_*
\end{equation}
which is the distance from the BH whereby the star's self-gravitational forces are overcome by the extreme tidal forces acting on it. 

Stars are swallowed by a non-spinning BH without tidal disruption if $r_{\text{t}} < r_{\text{g}} = 2GM_{\text{bh}}/c^2$, where $r_{\text{g}}$ is the BH's gravitational radius. This places an upper constraint on black hole mass for tearing a star apart (eg. \citealt{Hills1975,Rosswog2009}):
\begin{equation}\label{eqn:mbh_condition} 
M_{\text{bh}} < 2.5 \times 10^5 \left( \frac{R_{*}}{10^9\text{cm}} \right) ^{3/2} \left( \frac{M_{\text{*}}}{0.6 M_{\odot}} \right) ^{-1/2} M_{\odot}
\end{equation}
on condition of ignoring strong relativistic effects.

Following pericenter passage for single, highly disrupted encounters, approximately half of the material of the star is ejected from the system, whilst the other half remains bound to the BH in Keplerian orbits, subsequently falling back to the pericenter with fallback timescale \citep{EvansCandKochanek1989}:
\begin{equation}\label{eqn:t_fallback} 
t_{\text{fb}} \simeq \frac{r_{\text{p}}^3}{\sqrt{GM_{\text{bh}}}R_*^{3/2}}
\end{equation}
where $t_{\text{fb}}$ is dependent upon the energy distribution of the bound debris \citep{Rees1988,Lodato2009}. The fall back rate onto the black hole is:
\begin{equation}\label{eqn:m_fallback} 
\dot{M}_{\text{fb}} = \frac{1}{3} \frac{M_*}{t_{\text{fb}}} \left( \frac{t_{\text{fb}}}{t}\right) ^{5/3}
\end{equation}
and peaks at:
\begin{equation}\label{eqn:mdot_peak} 
\dot{M}_{\text{peak}} \approx \frac{1}{3 \delta^{5/3}} \frac{\sqrt{GM_{\text{bh}}}R_*^{3/2}}{r_{\text{p}}^3} M_*, 
\end{equation}
where the value of $\delta $ is dependent upon the type of star. For WDs, $\delta \approx 3.33$ for $M_{\text{wd}}>0.5M_{\odot}$ \citep{EvansCandKochanek1989}, and $\delta \approx 5.5$ for $M_{\text{wd}} \leq 0.5M_{\odot}$ \citep{Lodato2009}. $\dot{M}_{\text{peak}}$ occurs at a time:
\begin{equation}\label{eqn:t_peak} 
t_{\text{peak}} = \delta \frac{r_{\text{p}}^3}{\sqrt{GM_{\text{bh}}}R_*^{3/2}}
\end{equation}
after the disruption \citep{Shcherbakov2013}. The strength of the encounter is typically characterised by the parameter $\beta=r_{\text{t}}/r_{\text{p}}$.

The bound stellar debris circularizes around the black hole to form an accretion disc post-disruption. Early studies assumed that the accretion rate onto the BH, $\dot{M}_{\text{acc}}$, would trace $\dot{M}_{\text{fb}}$ if the viscous timescale, $t_{\text{visc}}$, is less than the fallback timescale, yet the physical justification for this link is becoming more uncertain \citep{Dai2013,Hayasaki2013,Hayasaki2016,Piran2015,Shiokawa2015,Bonnerot2016}. Significant deviation between $\dot{M}_{\mathrm{fb}}$ and $\dot{M}_{\mathrm{acc}}$ can occur if there is little energy dissipation  per orbit (low viscosity) and material is slow to circularise around the BH \citep{Ramirez-Ruiz2009}. This is more likely to occur in WTDEs whereby small self-intersection angles between debris streams (due to smaller general relativistic orbit precession effects), combined with the intersection point being approximately at the apogee, initially leads to low levels of energy dissipation per orbit and inefficient circularisation of debris around the BH \citep{Guillochon2015}. \citet{Macleod2016} estimate that this leads to a lowering of $\dot{M}_{\text{acc}}$ by a factor of $\sim$10 relative to $\dot{M}_{\text{fb}}$ (0.1 term in equation~(\ref{eqn:mdot_decay_viscous})). 

Initial mass fallback rates are likely to be highly super Eddington (eg. \citealt{Loeb1997,Macleod2016,Law-Smith2017}), such that the event is radiatively inefficient and its luminosity output Eddington-limited \citep{Haas2012,Khabibullin2014}, remaining approximately constant and only starting to decay once accretion rates become sub-Eddington. We note that the process of how falling-back debris is accreted onto the central object is still not well understood, especially for super-Eddington fallback rates. Furthermore, whilst super-Eddington accretion rates persist, the environment may favour jet production. If a jet is launched, it is likely to dominate the power output of the event if viewed on-axis, where the radiation will likely be non-thermal, whereas off-axis emission is likely to be thermal. There is also the possibility of radiatively-driven winds originating from the accretion disc, but we do not consider these here.

\begin{figure}
 \includegraphics[scale=0.95]{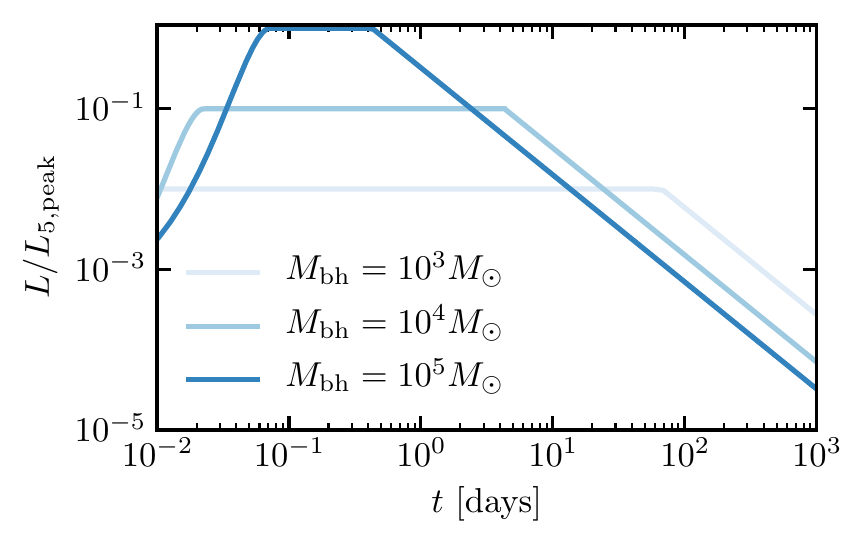} \label{fig:10$^3$}
 \caption{Simulated light curves for WTDEs based on the theory described in Section~\ref{sec:tdes}. Higher $M_{\mathrm{bh}}$ have faster rises to peak $L$, with fastest decay rates in lightcurve for $10^5M_{\odot}$ BHs. For comparison of peak luminosities of these events, we normalise each lightcurve by $L_{5\mathrm{, peak}}$, the Eddington luminosity of a $10^5M_{\odot}$ BH.}
 \label{fig:multi_wd_lcs}
\end{figure}

Following \cite{MacLeod2014}, if $t_{\text{visc}}>t_{\text{fb}}$, $\dot{M}_{\text{acc}}$ no longer traces $\dot{M}_{\text{fb}}$, with viscous expansion of the disc modifying accretion timescales according to \citep{Cannizzo2009,Cannizzo2011}:
\begin{equation}\label{eqn:mdot_decay_viscous} 
\dot{M}_{\text{acc}}(t)=0.1\dot{M}_{\text{peak}} \left( \frac{t}{t_{0}} \right) ^{-4/3},
\end{equation}
where $t_0 = 4/9 \alpha ^{-1}r_{\text{p}}^{3/2}(GM_{\text{bh}})^{-1/2}$ and $\alpha = 0.1$ for thick discs. The form of $\dot{M}_{\text{acc}}$ during its rise is currently not well constrained; we model it crudely here using a half-Gaussian centred on $t_{\text{peak}}$ with $\sigma =t_{\text{peak}}/4$, and normalise its peak to $L_{\text{Edd}}$. In this work, we assume that the observable X-ray light curve will trace $\dot{M}_{\text{acc}}$ and plot examples of our simulated lightcurves following the above description in Fig.~\ref{fig:multi_wd_lcs}.

The white dwarf experiences extreme compression perpendicular to the orbital plane, with this being maximised as it crosses pericenter. This is expected to trigger thermonuclear burning if its timescale 
is much shorter than the dynamical timescale (see eg. \citealt{LuminetJ1989,Brassart2008,Rosswog2009,Haas2012}). This burning originates in the tidal debris, with emission likely resembling type 1a SNe-like light curves in the optical band \citep{Rosswog2008,Rosswog2009,Macleod2016,Kawana2017}. Whilst thermonuclear burning may affect the amount of bound debris and subsequent accretion rate, no robust models of this currently exist and we neglect its impact in this work.

\subsubsection{Spectral properties}\label{sec:spectral_properties}
Following the modelling of emission from thick accretion discs (see \citealt{Balbus1998InstabilityDisks} for a review), the innermost sections of the disc, $R<5R_S$ (as in \citealt{Ulmer1999} and \citealt{Khabibullin2014}), radiate as a blackbody at temperature:
\begin{equation}\label{eqn:bb_spectrum} 
T_{\mathrm{bb}} \simeq  \left( \frac{L_{\mathrm{Edd}}}{4 \pi \sigma (5 R_{S})^2 } \right) ^{1/4} \, \text{K}
\end{equation}
where $L_{\mathrm{Edd}}$ is the Eddington luminosity of the BH. For IMBHs with masses of $10^3-10^5M_{\odot}$, this results in X-ray spectra with temperatures $\sim 0.2$keV. The soft emission from these events can be significantly affected by the interstellar medium's (ISM's) absorption along the line of sight \citep{Auchettl2017a}. For each TDE, we set the Galactic neutral hydrogen column density, $N_{\text{H}}$, to $5\times 10^{20}\text{cm}^{-2}$, close to the estimated extra-galactic median of $N_{\text{H}}$ \citep{Esquej2008}. We model the WTDE's spectrum using the XSPEC model \texttt{tbabs(zbbody)}, but neglect modelling any spectral evolution. 

An optically-thick extended envelope may form from stellar debris around the black hole, reprocessing a fraction of the X-ray accretion flare into the UV and optical bands (eg. \citealt{Loeb1997,Ulmer1998}), with the extent of reprocessing potentially viewing angle dependent \citep{Roth2016,Dai2018}. As we are interested in exploring eROSITA's sensitivity to the X-ray signatures of WTDEs, we keep to this simplified model of emission from the event and ignore reprocessing in this work (further discussion in section~\ref{sec:caveats}).

\subsection{AGN population}
\label{sec:agn_population}
We construct a realistic AGN population based on Section 2 of \citet{Clerc2018SyntheticSimulations}. AGN are generated down to several orders of magnitude below the eROSITA flux detection limit for point sources at the end of the four year all-sky survey. To lower SIXTE's computational expense, we then split the generated AGN \texttt{simput} file into two parts by selecting those with flux above and below $f_{\textrm{split}}=10^{-16}\text{ergs}^{-1}\text{cm}^{-2}$ in the 0.5-2 keV band respectively. AGN with flux above $f_{\textrm{split}}$ are stored in their own \texttt{simput} file. For AGN with flux below $f_{\textrm{split}}$, we stack the spectra of all sources below this limit and assign this spectrum to a new, artificial extended source of same dimensions as the original AGN population for the sky patch. This extended source is then placed into a separate \texttt{simput} file and models the cosmic X-ray background (CXB) for simulations. Whilst flaring AGNs may represent a large fraction of detected transient sources in eRASS, we do not model AGN time variability in this work and discuss this further in Section~\ref{sec:sim_erosita_observations}.  

\subsection{Soft X-ray background}
\label{sec:xrb}
For each sky patch simulated, we determine the diffuse X-ray background due to the SXB and CXB at that position using the {\it HEASARC Soft X-Ray Background Tool}\footnote{\url{https://heasarc.gsfc.nasa.gov/cgi-bin/Tools/xraybg/xraybg.pl}} to obtain the integrated flux and spectrum for the associated sky position. Using XSPEC, we fix $N_{\text{H}}$ and fit the spectrum with the model: \texttt{apec+wabs(apec+powerlaw)}, obtaining a set of best fitting parameters for each component (\texttt{apec} is a model of the X-ray emission of collisionally-ionized diffuse gas, whilst \texttt{powerlaw} is an approximate model of the CXB at the given sky position). We then use the best fitting parameter values to construct a \texttt{simput} file for the SXB using \texttt{apec+wabs(apec)}. Information about the {\it powerlaw} parameters is discarded as we model the CXB using the stacked AGN below the flux detection limit as described in Section~\ref{sec:agn_population}. 

\section{\lowercase{e}ROSITA detection sensitivity} \label{sec:detection_sensitivity}
The standard processing of the eROSITA all-sky survey data will use a tiling system that divides the sky into 3.6$\deg \times$3.6$\deg$ sky fields. The exposure of each tile over an eRASS is dependent upon its sky position; those near the ecliptic equator have short exposures and are visited only over a single Earth day, whereas those at the poles are visited over consecutive Earth days and have longer exposures. Polar fields therefore present the opportunity to discover lower flux transients, but cover a smaller fraction of the sky. 

Given the observational signatures of WTDEs described in Section~\ref{sec:tdes} and for different sky positions, we are interested in exploring the detection sensitivity of eROSITA to different WTDE parameter configurations. To do this for every eROSITA skyfield would be too computationally expensive; we instead look at this across two different skyfields labelled \textit{2090} (equatorial) and \textit{110135} (intermediate), with exposure times of $\sim$0.2ks and $\sim$0.5ks respectively in eRASS1\footnote{Successive eRASS scans will also have similar exposures, but we only consider eRASS1 in this work.}.

\subsection{Grid simulations}
Each WTDE can be parametrised in terms of $(M_{\text{bh}}, M_{\text{wd}}, \beta, z, \mathcal{T})$, where $\mathcal{T}=t_{\text{first}}-t_{\text{flare}}$, with $t_{\text{first}}$ being the time when the sky position of that event first enters eROSITA's FOV during eRASS1 and $t_{\text{flare}}$ is the time of the disruption. However, the probability of a given event being detected is most dependent on the subset $(M_{\text{bh}}, z, \mathcal{T})$. This is due to higher $M_{\text{bh}}$ values leading to faster decay of $\dot{M}_{\text{acc}}$ and greater peak luminosities since $L_{\text{Edd}} \propto M_{\text{bh}}$; 
whilst as $z$ increases, the measured flux in the soft X-ray regime drops due to increasing distance to the event and redshifting of the intrinsically soft spectrum. 
Positive $\mathcal{T}$ values have the TDE flare occurring before they first enter the FOV (and vice versa) -- so as $\mathcal{T}$ increases, the flux of the source when first observed decreases. We note that as some skyfields are visited over multiple days during an eRASS, a negative $\mathcal{T}$ does not prevent detection.  $M_{\text{wd}}$ only ranges between 0.2-1.4$M_{\odot}$ and modifies $\dot{M}_{\text{peak}}\propto M_{\text{wd}}$, thus has a smaller impact on the detection efficiency, $\mathcal{D}$, in comparison.

We construct a grid of WTDEs in each skyfield, with spacings between events of 0.8\degree ~in RA and Dec. For each skyfield, we consider five different $M_{\text{bh}}$ values: $10^3$, $10^{3.5}$, $10^4$, $10^{4.5}$ and $10^5M_{\odot}$. $M_{\text{bh}}$ is fixed within each grid, and we consider a low and high $z$ set of WTDEs in each skyfield where $z$ increases uniformly between 0.01-0.16 and 0.17-0.32 respectively. For each $M_{\text{bh}}$ grid, we vary $\mathcal{T}$ of all WTDEs according to: [-0.75, -0.25, 0, 0.25, 1, 5, 10, 20, 45, 90]d and [-5, -2.5, -1, 0, 1, 5, 10, 20, 45, 90]d for the 2090 and 110135 skyfields respectively. For all simulations, we set $M_{\text{wd}}=0.5M_{\odot}$, close to the measured mean mass from SDSS observations \citep{Kepler2007}, but also sufficiently small so that WDs are disrupted instead of being swallowed whole for all simulated $M_{\mathrm{bh}}$ values (ie. the condition in equation~\ref{eqn:mbh_condition} remains valid). We further simplify modelling via fixing $\beta = 1$ for all disruptions, since encounters where $r_{\mathrm{p}}\approx r_{\mathrm{t}}$ are anticipated to be most common (eg. \citealt{Rees1988,Guillochon2015,Stone2016}). 

This leads to 200 WTDE grids in total\footnote{ie. (5 different $M_{\mathrm{bh}}$) $\times$ (2 $z$ ranges) $\times$ (10 $\mathcal{T}$ values) for each skyfield}, and we then run 10 differently seeded SIXTE simulations of each one over eRASS1. The resulting event files are merged with the simulated event files of the AGN and SXB over eRASS1 (sources in the AGN \texttt{simput} are shuffled in position for each seeded run and then simulated). Subsequent source detection and characterisation is performed on the merged event files using the most recent release of the eROSITA analysis software, \textit{eSASS}. Using the software \textit{stilts}\footnote{\url{http://www.star.bris.ac.uk/~mbt/stilts/}}, we then match the input population of TDEs to the list of eSASS detected sources based on the position estimates. To reduce the number of false matches between input population and detected AGN, we also impose the constraint that eSASS estimated the detected, matched source to have $<30\%$ and $<5\%$ of their photon counts between 2-5keV and above 5keV respectively.

\subsection{Detection efficiency}
\begin{figure*}
 \includegraphics[scale=0.85]{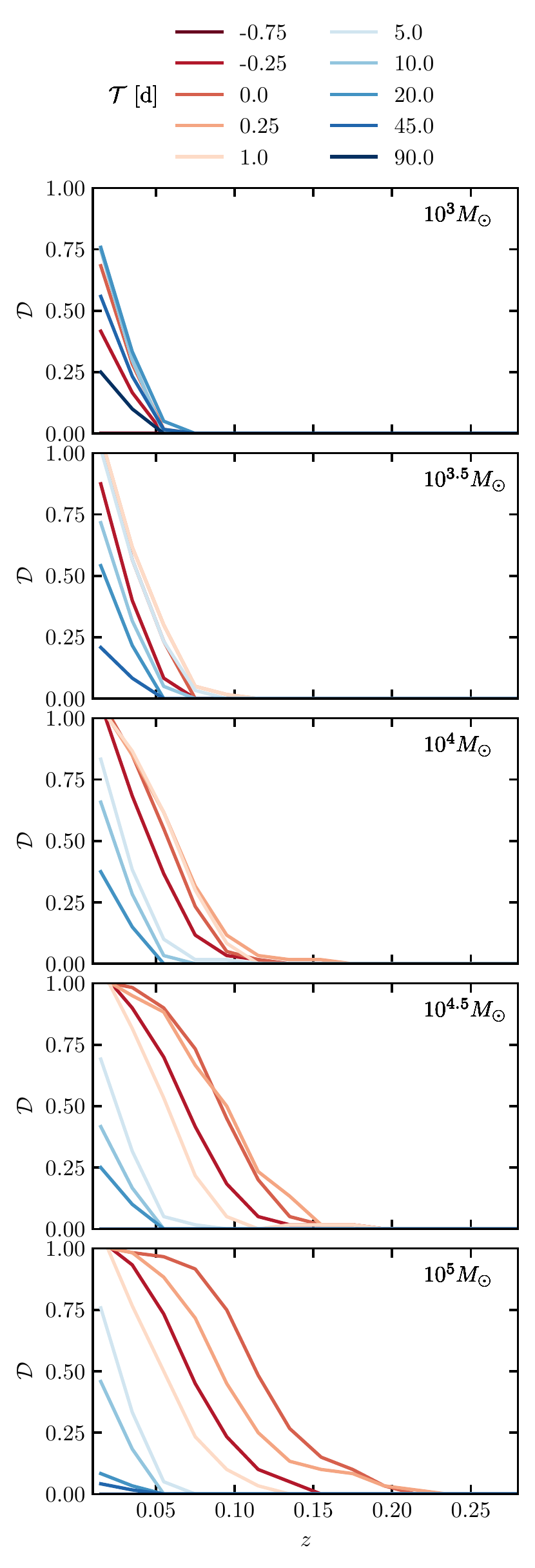}
 \includegraphics[scale=0.85]{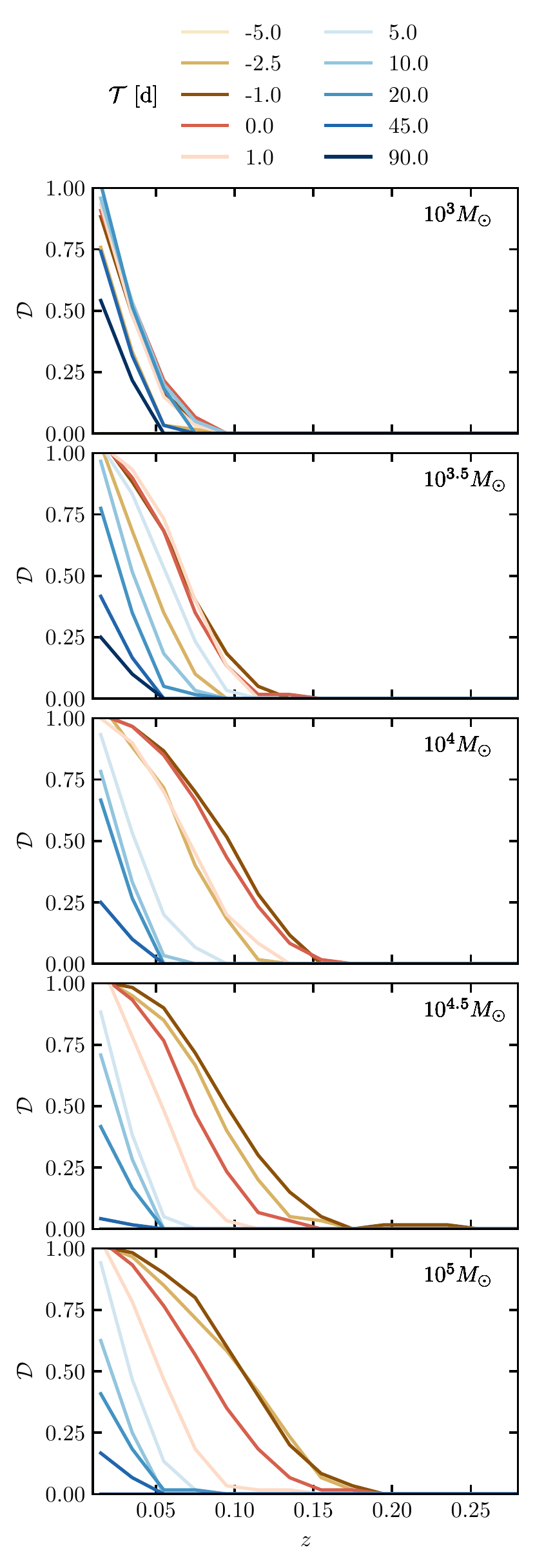}
 \label{fig:det_efficiency}
 \caption{Estimated detection efficiency, $\mathcal{D}$, of WTDEs as function of $z$ for different IMBH masses (top-to-bottom) and different $\mathcal{T}$ (note the different color
coding). Left and right columns show the equatorial and intermediate skyfields, respectively.}
\end{figure*}

$\mathcal{D}(M_{\mathrm{bh}}, \mathcal{T}, z)$ for each configuration of parameters is estimated by computing the fraction of times each WTDE was detected by eSASS across the 10 seeded runs (Fig.~\ref{fig:det_efficiency}). To reduce Poisson noise in estimated $\mathcal{D}(M_{\mathrm{bh}}, \mathcal{T}, z)$ due to running only 10 seeded simulations, we smooth $\mathcal{D}$ by applying a Savitzky-Golay filter\footnote{using the \texttt{scipy.signal.savgol\_filter} implementation.} \citep{Savitzky64} to each curve, with window length 3 and degree of polynomial of 1.

From Fig.~\ref{fig:det_efficiency}, we see that for both skyfields, events involving higher black hole masses will be detectable out to a larger maximum redshift, $z_{\text{max}}$, relative to lower $M_{\mathrm{bh}}$. Under these conditions, eROSITA probes each event when $L \approx L_{\mathrm{Edd}}\propto M_{\mathrm{bh}}$; either due to $\dot{M}_{\mathrm{acc}}$ being capped at $\dot{M}_{\mathrm{Edd}}$, or $L$ having decayed insignificantly since flaring. This is in contrast to the detectability of TDEs from MS stars with eROSITA, where \citet{Khabibullin2014} estimate $z_{\text{max}}$ to be independent of $M_{\mathrm{bh}}$. This
likely stems from the larger decay timescales of MS TDEs (months to years) causing smaller drops in luminosity as a function of $\mathcal{T}$, such that on average they should be observed with eROSITA much closer to their $L_{\text{Edd}}$.
The blackbody temperature $T_{\mathrm{bb}}\propto M_{\mathrm{bh}}^{-1/4}$ (section~\ref{sec:spectral_properties}) for WTDEs has relatively little impact on $\mathcal{D}$, in comparison to $\mathcal{T}$ or $z$. 

The longer exposure times of intermediate\footnote{These fields will have deeper exposure times as a result of multiple visits of the skyfield over consecutive days, as opposed to only a single day.} relative to equatorial skyfields generally boost $\mathcal{D}$ for most $M_{\mathrm{bh}}$, though the magnitude of this boost is dependent on $\mathcal{T}$ and $M_{\mathrm{bh}}$ (Fig.~\ref{fig:det_efficiency}). For the general case of a constant luminosity source, $\mathcal{D}$ should always be greater for an intermediate compared with equatorial sky field if running detection over the stacked events of an eRASS. However for faster transient sources, Fig.~\ref{fig:det_efficiency} highlights a more subtle feature of eROSITA's transient detection ability -- $\mathcal{D}$ is determined by a trade-off between exposure, average effective area over this exposure, and the luminosity and decay rate of the transient. Across all skyfields, those with longer exposures see a decrease in the rate at which a source located in one moves through eROSITA's FOV. For example, an equatorial field has 6 consecutive visits over an Earth day every 4 hours, whereas an intermediate field has the same frequency of visits but over a $\sim 5$ day period. The effective area the source is observed with peaks during the middle of this period, when the source passes through the centre of eROSITA's FOV, and decreases either side of this. This means that on the first eROSITA day a source is observed, the average effective area would be greater for the equatorial skyfield relative to the intermediate one, so a source of a given luminosity would have a greater chance of being detected in the equatorial field in this case. If one considers fast decaying transients (such as for the $M_{\text{bh}}=10^5M_{\odot}$ WTDEs), the same effect will be present. However in this case, the brightest stage of a WTDE could be observed on average with a lower effective area for the intermediate field compared to an equatorial field. With the source flux decaying so rapidly for these events, by the time it later gets observed in an intermediate field at a higher effective area, its flux has decreased so greatly that the added exposure time has little impact on increasing $\mathcal{D}$. This effect contributes to the observed behaviour of $\mathcal{D}$ curves in Fig.~\ref{fig:det_efficiency}, and will be more noticeable the faster the transient decays (ie. for the more massive IMBHs) and around $\mathcal{T}\sim 0$d. 

Regardless of skyfield (and as expected), our best chance for detection of a WTDE is catching the event close to its flaring, with detection sensitivity rapidly dropping for higher $M_{\mathrm{bh}}$ in the following days. For instance, between $\mathcal{T}=1$d and $\mathcal{T}=5$d for the equatorial skyfield, the redshift at which we detect 50\% of events drops from 0.07 to 0.04 for $M_{\mathrm{bh}}=10^5M_{\odot}$, but is roughly constant during this time interval at $z\approx$0.04 for $M_{\mathrm{bh}}=10^3M_{\odot}$. For $\mathcal{T}>5$d, we are only able to detect events up to $z\sim 0.05$ for all IMBHs. Negative $\mathcal{T}$ values see the WTDE flaring after eROSITA first visits it during an eRASS, with intermediate-like fields presenting the best opportunity to detect negative $\mathcal{T}$ cases. For these, eROSITA will catch the event flaring in the middle of a multi-day visit and thus offers the potential to sample the rise, peak, and decay of its lightcurve.

\section{Estimate of rate}\label{sec:rate_estimate}
In this section, we will first derive an updated estimate on the rate density of WTDEs in the local universe, before using this along with the inferred detection sensitivities to estimate eROSITA's detection rate of these events.

\subsection{Intrinsic rate of WTDEs} \label{sec:intrinsic_rate}
Following \citet{Shcherbakov2013}, we assume that the IMBH population in the local universe mainly resides in globular clusters (GCs) and dwarf galaxies (DGs), with these two being the main contributors to the WTDE rate density\footnote{the number of WTDEs per year per Gpc$^3$}, $\mathcal{R}$. Furthermore, we assume that GCs host BHs with $10^3M_{\odot}<M_{\mathrm{bh}}<10^4M_{\odot}$, and DGs those with $M_{\mathrm{bh}}>10^4M_{\odot}$. 

The contribution to $\mathcal{R}$ from GCs is estimated based on \citet{Fragione2018}, which uses a semi-analytic model to evolve a primordial population of globular clusters within massive host galaxies over cosmic time. The evolution takes into account mass loss via stellar winds, the loss of stars through evaporation, two body interactions and tidal stripping by the host galaxy. Furthermore, they also include interactions between the IMBH and stellar mass black holes that can eject the IMBH from its GC due to recoil from asymmetric gravitational wave emission. Combined with estimates of the local number density of globular clusters in \citet{Rodriguez2015}, they present estimates of the WTDE rate density in the local universe from elliptical and spiral galaxies for three different $M_\mathrm{*}$. 

\begin{table}
	\centering
	\caption{Estimated rates of WTDEs due to elliptical and spiral galaxies computed from the mean of estimated values up to $z=0.24$ from Fig. 9 of \citet{Fragione2018}.}
	\label{tab:fragione2018}
	\begin{tabular}{lccr} 
		\hline
		 $M_*$ ($M_\odot$) & $\mathcal{R}_{\mathrm{e}}(M_*)$ (Gpc$^{-3}$yr$^{-1}$) & $\mathcal{R}_{\mathrm{s}}(M_*)$ (Gpc$^{-3}$yr$^{-1}$)\\
		\hline
		$10^{10}$ & 18 & 14 \\
		$5\times10^{10}$ & 6 & 4\\
		$10^{11}$ & 7 & 3 \\
		\hline
	\end{tabular}
\end{table}

To estimate the total rate density from the estimates for specific host galaxy masses in table \ref{tab:fragione2018}, we sum the weighted mean contribution from elliptical and spiral hosts:

\begin{equation} \label{eqn:rate_gc}
\mathcal{R}_{\text{GC}}=\frac{\sum_i \Phi (M_{\mathrm{*},i}) \mathcal{R}_{\mathrm{e}}(M_{*,i})}{\sum_i \Phi (M_{*,i})}+\frac{\sum_i \Phi (M_{*,i}) \mathcal{R}_{\mathrm{s}}(M_{*,i})}{\sum_i \Phi (M_{*,i})}
\end{equation}
where $\Phi (M_{*,i})$ are computed using the Schechter function \citep{Schechter1976AnGalaxies.}:
\begin{equation}
\Phi (M_*)=\Phi _{\mathrm{c}} \left( \frac{M_\mathrm{*}}{M_\mathrm{c}}\right) ^{-\alpha _\mathrm{c}}e^{-M_*/M_\mathrm{c}},
\end{equation}
with $\Phi _c$ a normalising constant, $M_\mathrm{c}=10^{11.14}M_{\odot}$ and $\alpha _\mathrm{c}=1.43$, using parameter estimates from EAGLE cosmological simulations \citep{Furlong2015EvolutionSimulations}. Combining equation~\ref{eqn:rate_gc} and data from table~\ref{tab:fragione2018} yields a local rate density from GCs of $\sim30$~Gpc$^{-3}$yr$^{-1}$ (17 and 13~Gpc$^{-3}$yr$^{-1}$ from elliptical and spiral galaxies respectively).

We then assume the dwarf galaxy contribution to $\mathcal{R}$ has the form:
\begin{equation} \label{eqn:rate_dg}
\mathcal{R}_{\text{dg}} = f_{\text{oc}} n_{\text{dg}} \dot{N}_{\text{dg}} 
\end{equation}
where $f_{\text{oc}} \sim 0.1$ is the estimated occupation fraction in DGs of BHs with $M_{\text{bh}}<10^5M_{\odot}$ \citep{AlvesBatista2017}, although we note that \cite{Miller2015X-RAYFRACTION} estimate a lower limit on $f_{\text{oc}}$ at 95\% confidence to be $\sim$0.2 for galaxies with $M_*<10^{10}M_{\odot}$. Our resulting estimated WTDE detection rates will likely be conservative with respect to $f_{\text{oc}}$. $\dot{N}_{\mathrm{dg}}\sim 10^{-6}\, \mathrm{yr}^{-1}$ is the estimated rate of WTDEs per IMBH computed in \citet{MacLeod2014} and $n_{\text{dg}}$ is the local number density of dwarf galaxies. The latter is estimated via integration of a double Schechter function with best fitting parameters from \citet{Blanton}, obtained via their fitting of the luminosity function of extremely low luminosity galaxies observed with the SDSS (r-band), corrected for selection effects. Over the range of this integral (bounds defined below), it is assumed that only dwarf galaxies contribute to the number density. 

There is currently no accepted, clear distinction in the literature between dwarf and non-dwarf galaxies. However for the purpose of estimating $n_{\text{dg}}$, this is only a minor issue since $n_{\text{dg}}$ is far less sensitive to the choice of upper bound compared with the lower. A coarse cut-off has previously been considered at $M_{r}\approx-18$ (also approximately the magnitude of the Large Magellanic Cloud; \citealt{McConnachie2012}), we choose this as the upper bound of this integral. 
We then consider two different lower bounds herein. The first is at $M_r = -6$, which represents an extrapolation of the best fit Schechter function in \citet{Blanton} down to the lowest luminosity dwarf galaxies observed outside the Milky Way (see for example \citealt{McConnachie2012}). This also roughly corresponds to requiring a minimum of $\sim 10^6M_{\odot}$ contained within the half-light radius of the DG, based on Fig. 11 of \citet{Torrealba2018}. The second choice, at $M_r = -12$, is a more conservative estimate, which is approximately the lowest value of the $M_r$ data used for model fitting in \citet{Blanton}. These two different lower bounds yield estimated $n_{\text{dg}}$ of 2.6 and 0.3$h^3\text{Mpc}^{-3}$ respectively, with the former being approximately an order of magnitude greater than a previous lower limit for $n_{\text{dg}}$ of 0.12$h^3\text{Mpc}^{-3}$ estimated in \citet{Loveday1997TheGalaxies}. However, the authors of that work suggest that their reported $n_{\text{dg}}$ is likely to be significantly lower than the true value due to the various assumptions they make on galaxy clustering that lead to underestimation of $n_{\text{dg}}$. 

Adopting $h=0.678$ from \citet{Planck2015}, this leads to our rate density estimates from BHs in DGs with $10^4M_{\odot} < M_{\mathrm{bh}} < 10^5M_{\odot}$ of:
\[
  \mathcal{R}_{\text{dg}} \approx
  \begin{cases}
    10 \, \text{Gpc}^{-3}\text{yr}^{-1} & \text{if } n_{\mathrm{dg}}=0.3h^3\text{Mpc}^{-3} , f_{\text{oc}}=0.1 \\
    80 \, \text{Gpc}^{-3}\text{yr}^{-1} & \text{if } n_{\mathrm{dg}}=2.6h^3\text{Mpc}^{-3}, f_{\text{oc}}=0.1.
  \end{cases}
\]
The DG contribution to $\mathcal{R}_{\text{dg}}$ is expressed conditional on our adopted $n_{\text{dg}}$ and $f_{\text{oc}}$ values; this highlights that if, for example, $f_{\text{occ}}$ were to double, then we would expect that $\mathcal{R}_{\text{dg}}$ would rise to 160$\text{ Gpc}^{-3}\text{yr}^{-1}$ for the case where $n_{\mathrm{dg}}=2.6h^3\text{Mpc}^{-3}$.

Similar to previous studies, we find that dwarf galaxies are the dominant source of WTDEs, although even our highest $\mathcal{R}_{\mathrm{dg}}$ estimate is roughly a factor of 20 lower than in \citet{Shcherbakov2013}. The main cause of this disagreement is that we use $\dot{N}_{\mathrm{dg}}\sim 10^{-6}\, \mathrm{yr}^{-1}$, compared with their $0.15 \times 10^{-3}\, \mathrm{yr}^{-1}$ (they assume a rate of stellar tidal disruption of $10^{-3}\, \mathrm{yr}^{-1}$, and 15\% of these events involve white dwarfs) which bears the problem of BH overgrowth; the IMBHs would rapidly grow above $10^5M_{\odot}$ via WTDEs alone if those rates were sustained. Furthermore, they assume that all dwarf galaxies will host an IMBH with mass low enough for a valid WTDE, whereas we estimate 10\% of BHs in DGs will have $M_{\mathrm{bh}}<10^5M_{\odot}$.

\subsection{eROSITA detection rate}
For eRASS1, we consider the set of WTDEs that flare with $-90\mathrm{d}<t_{\text{flare}}<180\mathrm{d}$ (relative to start of eRASS1), and within the spherical volume enclosed by $z < 0.24$, as potentially detectable events. This range is chosen based on the $\mathcal{D}$ curves in Fig.~\ref{fig:det_efficiency}, where $\mathcal{D}(M_{\mathrm{bh}}, \mathcal{T}>90\mathrm{d}, z>0.24) \sim 0$. Using our WTDE rate density estimates ($\mathcal{R}$), this volume and period corresponds to $N_{\mathrm{tot, gc}}=180$ and $N_{\mathrm{tot, dg}}=470$ (40) events from globular clusters and dwarf galaxies respectively (bracketed number for the case of $n_{\text{dg}}=0.3 h^3\text{Mpc}^{-3}$). The number of WTDE detections during each eRASS is then estimated by drawing  $N_{\mathrm{tot, gc}}$ and $N_{\mathrm{tot, dg}}$ random WTDE configurations and computing the number of these that are detected (further described below). We then repeat this 1000 times, replicating 1000 eRASS scans of each host class, to study the distribution of the expected number of detections.

We consider the sky to be divided into two parts which have exposure above and below the $\sim$0.5ks exposure of an intermediate skyfield, occupying 90\% and 10\% of the sky each. These weights are obtained from the estimated exposure map predicted for eROSITA's All-Sky Survey (J. Robrade, private communication). eROSITA's detection sensitivity to WTDEs in each region is then modelled using the estimated $\mathcal{D}$ for the 2090 and 110135 skyfields respectively. The number of detections in eRASS1 from globular clusters, $N_{\text{gc}}$, is estimated via:
\begin{equation}\label{eqn:total_rate_gc}
N_{\mathrm{gc}}=N_{\text{gc, eq.}}+N_{\text{gc, int.}} 
\end{equation}
where $N_{\text{gc, eq.}}$ ($N_{\text{gc, int.}}$)  is 90\% (10\%) of the estimated number of detections over the whole sky when assuming a detection sensitivity of the equatorial (intermediate) field. Similarly, the number of WTDE detections from dwarf galaxies is:
\begin{equation}\label{eqn:total_rate_dg}
N_{\mathrm{dg}}=N_{\text{dg, eq.}}+N_{\text{dg, int.}},
\end{equation}
where definitions follow those for equation~\ref{eqn:total_rate_gc}.

Our simulations yield $\mathcal{D}(M_{\mathrm{bh}}, \mathcal{T}, z)$ in each skyfield. To cast this into a form easier to estimate the rate of WTDE detection, we infer eROSITA's detection sensitivity to all WTDEs occurring up to $z_{\mathrm{max}}$ as a function of $\mathcal{T}$, via marginalising over $z$:

\begin{equation}\label{eqn:z_marginalised}
\mathcal{D}(M_{\mathrm{bh}}, \mathcal{T}) = \int \mathcal{D}(M_{\mathrm{bh}}, \mathcal{T}, z)p(z)\mathrm{d}z
\end{equation}
where $p(z)$ is the expected $z$ distribution of WTDEs. To estimate this (due to lack of observational evidence of WTDEs), we split up the local universe into a series of nested spherical shells and assume no significant evolution in the TDE hosts within this volume. Setting each shell to have the same event number density, $n$, in comoving space, then the number of events per shell, d$N$, is expected to be:
\begin{equation}
\mathrm{d}N = n_{\text{}}\mathrm{d}V_{\text{sh}}(z)
\end{equation}
where $\mathrm{d}V_{\text{sh}}(z)$ is the comoving volume of the shell, with probability of finding an event at redshift z being:
\begin{equation}
p(z)\approx \frac{V_{\text{sh}}(z)}{V_{\text{max}}} \sim \frac{4\pi z^2}{\frac{4}{3}\pi z_{\text{max}}^3}= Az^2.
\end{equation}
with $A$ is defined such that $\int p(z) \mathrm{d}z =1$. 

After marginalisation, only $M_{\text{bh}}$ and $\mathcal{T}$ require drawing to evaluate $\mathcal{D}$ for a particular WTDE configuration. $M_{\text{bh}}$ is randomly chosen from $[10^{3}$, $10^{3.5}$, $10^{4}] M_{\odot}$ for globular clusters, and $[10^{4}$, $10^{4.5}$, $10^{5}] M_{\odot}$ for dwarf galaxies (effectively crudely sampled from a log-uniform distribution). A $\mathcal{T}$ sample for each WTDE is obtained by combining randomly drawn $t_{\text{flare}}$ and $t_{\text{first}}$ values, where $t_{\text{flare}}$ is drawn from $\mathcal{U}(-90\mathrm{d}, 180\mathrm{d})$. For drawing samples of $t_{\text{first}}$, we generate a random sky position and compute the time a source located there first enters eROSITA's FoV\footnote{1.02 $\deg$ diameter} in eRASS1 (using the SIXTE task \texttt{ero\_vis} and a spacecraft attitude file). Random source positions are drawn by assuming WTDEs are equally likely to occur at all points on the sky, and are drawn randomly from the surface of a sphere via normalising a set of three random numbers generated from $\mathcal{U}~(0, 1)$ \citep{Marsaglia1972}. Finally, we draw $M_{\text{bh}}$ from the range $10^3-10^4M_{\odot}$ and $10^4-10^5M_{\odot}$ for WTDEs in globular clusters and dwarf galaxies, respectively. 

\begin{figure}
 \includegraphics[scale=0.9]{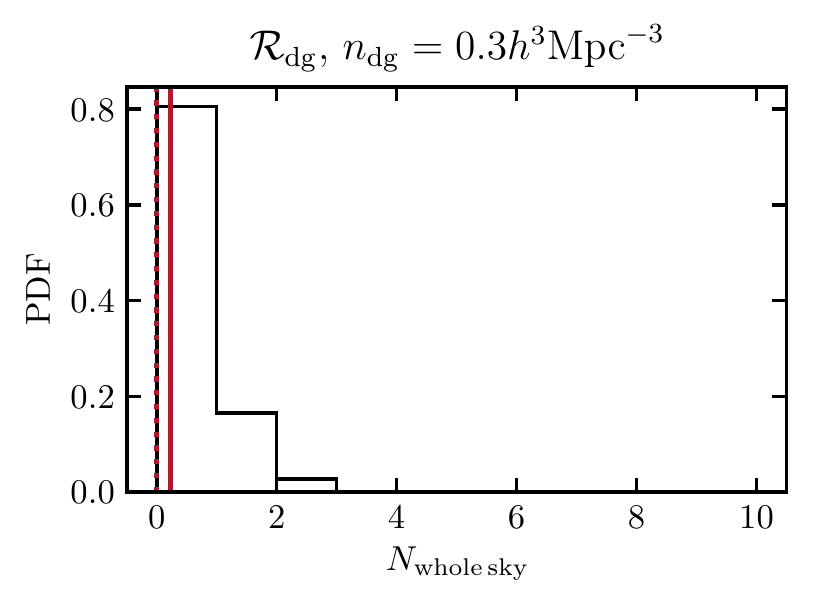}
 \includegraphics[scale=0.9]{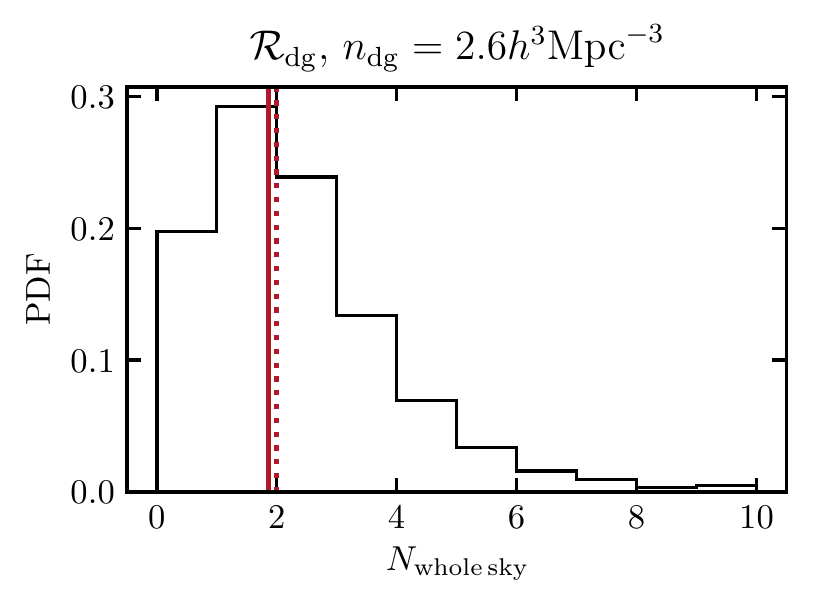}
  \includegraphics[scale=0.9]{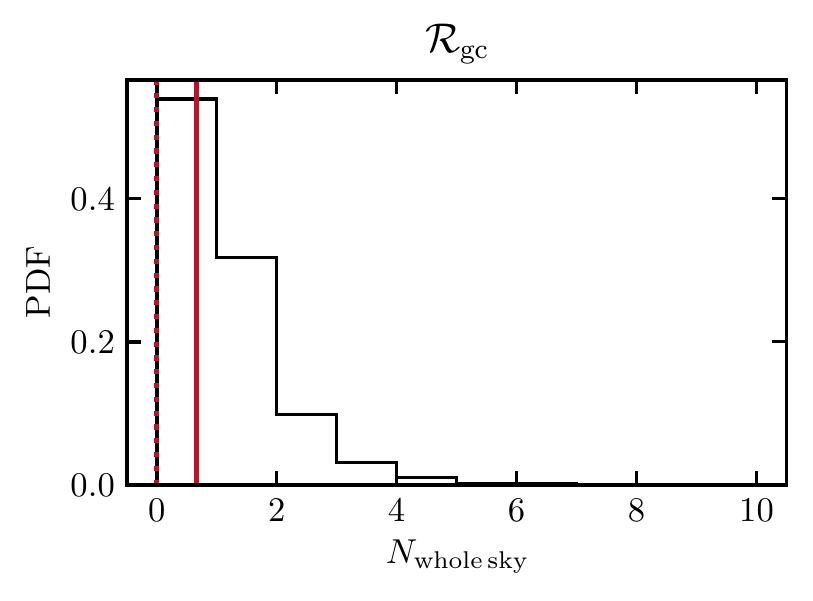}
 \caption{Distributions of $N_{\text{whole sky}}$ -- the expected number of detections up to eRASS8 over the whole sky, from dwarf galaxies (top two panels) and globular clusters (bottom). Dotted and solid vertical lines mark the median and mean of samples from the distribution.} \label{fig:n_detection_histograms}
\end{figure}

For each drawn $(M_{\text{bh}},\, \mathcal{T})$, we estimate the probability of detecting its associated WTDE, $\mathcal{D}$, via linear interpolation of equation~\ref{eqn:z_marginalised}. We generate a random number, $x$, from $\mathcal{U}(0, 1)$. If $x<\mathcal{D}$, then we classify the event as detected; otherwise it is a non-detection. We do this for each of the $N_{\text{tot,gc}}$ and $N_{\text{tot,dg}}$ detectable WTDEs in an eRASS, and repeat this process 1000 times. From this, we construct estimates for the number of WTDE detections per eRASS for the globular cluster and dwarf galaxy populations, with the distributions for the number of detections per eRASS presented in Fig.~\ref{fig:n_detection_histograms}. 

Scaling estimates for eRASS1 up to eRASS8, we estimate that over its 4 year all-sky survey, eROSITA may detect $\sim$2 WTDEs from dwarf galaxies (for $n_{\text{dg}}=2.6h^3\text{Mpc}^{-3}$) and $\sim$1 from globular clusters. On the other hand, if the more conservative estimate for $n_{\text{dg}}$ is more accurate, then we expect no detections from dwarf galaxies.

\begin{figure}
 \includegraphics[scale=0.9]{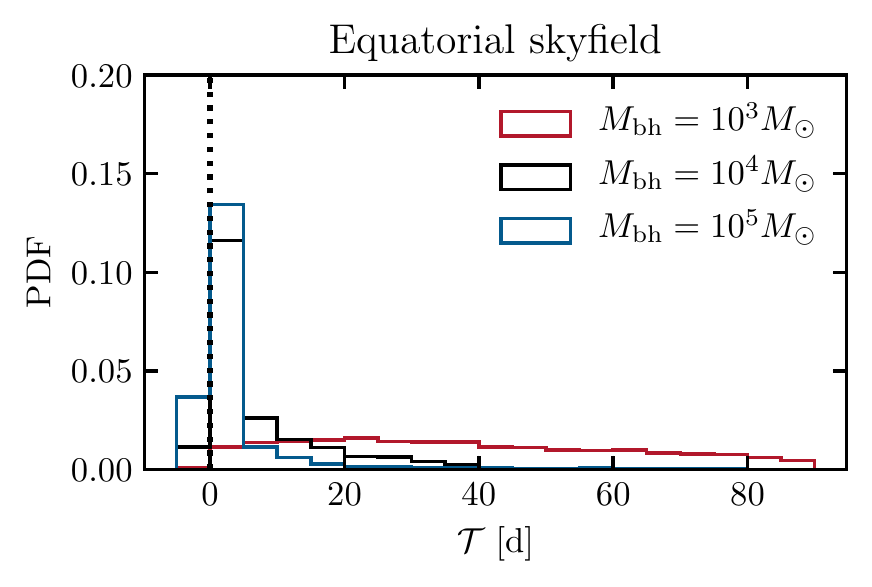}
 \includegraphics[scale=0.9]{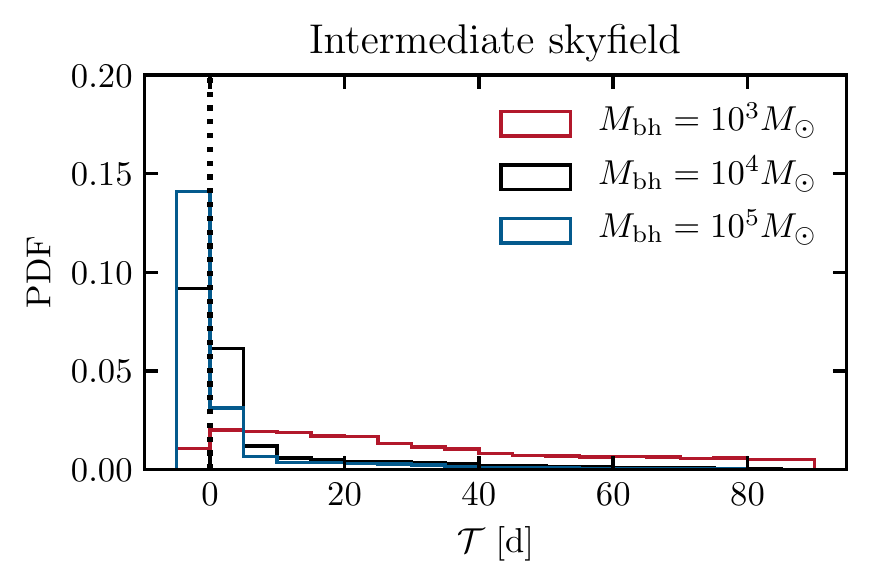}
 \caption{Inferred $\mathcal{T}$ distribution for detected WTDEs for different black hole masses, with top and bottom panels for the equatorial and intermediate skyfields respectively. Events detected with $\mathcal{T}<0$ (left of the black dotted vertical line) will flare after eROSITA has visited it for the first time during an eRASS. For the deeper exposure fields, the source will be scanned multiple times over consecutive days. In agreement with inferences drawn from Fig.~\ref{fig:det_efficiency}, as $M_{\mathrm{bh}}$ increases, $\mathcal{D}$ drops off more rapidly with $\mathcal{T}$.}\label{fig:toffset_histograms}
\end{figure}

As a byproduct of this set of Monte Carlo simulations, we can also infer the $\mathcal{T}$ distributions of WTDEs classified as detected for different black hole masses and different sky positions (Fig.~\ref{fig:toffset_histograms}). For $M_{\mathrm{bh}}>10^4M_\odot$, nearly all detected events are within 5 days of their flaring, whereas $M_{\mathrm{bh}}<10^4M_\odot$ events have much broader $\mathcal{T}$ distributions. Transitioning to fields with deeper exposure shifts the peak of the distribution towards lower $\mathcal{T}$. 

\section{Discussion} \label{sec:discussion}
 We have simulated eROSITA observations of WTDEs and estimated the number of detections during eROSITA's four year all-sky survey. However, a detection is different to identification -- for the latter, we are confident in our classification of the object type. Prior to building up a sample of observed WTDEs, one must be able to pick out and identify the WTDEs hidden in eROSITA's source catalogues. In the following section, we first consider predicted multi-wavelength signatures for these events that could be used to assist mining the large eROSITA datasets. We then briefly discuss the caveats of our assumptions made in inferring eROSITA's detection efficiency that may alter our detection rate estimates. 

\subsection{Identification and multi-$\lambda$ signatures}
WTDEs will be non-detectable at X-ray wavelengths in the subsequent eRASS sky pass 6 months after first detection (eg. Fig.~\ref{fig:det_efficiency}) -- for most eROSITA detections we may only have an X-ray lightcurve constructed from its $\approx$40s visits. From this sampling alone, it will be challenging to accurately classify a transient as a WTDE, or identify a candidate which requires further multi-wavelength followup. 
To overcome this and greatly reduce the number of false-positive WTDE identifications, it is vital to consider potential observational signatures of their hosts that already exist in the wealth of multi-wavelength sky surveys available. 

The majority of new X-ray variable sources detected per eRASS will likely be flaring stars and AGN. To reduce this set of newly detected flaring sources to a smaller group of WTDE candidates, an initial cross-match with multi-wavelength surveys should be performed. For instance, one could discard transients associated with centres of known massive galaxies as their BHs will exceed the mass
limit for successfully disrupting a WD. Similarly, stellar counterparts with significant proper motion measured by \textit{Gaia} could be rejected. We note that based on the extent of theoretical uncertainties and lack of significant observational evidence surrounding WTDEs, we anticipate selection of these events will be particularly difficult during eRASS1. However by eRASS2, we will have a much better understanding of the variable X-ray sky, thus allowing for a cleaner selection of the new flaring sources in following eRASS. Of the new X-ray transients that remain, additional information will be needed to support classification of the event as a WTDE. For instance, optical follow-up of a candidate may allow us to ascertain the nature of its host or detect an associated optical transient, whilst obtaining an estimated $z$ for the event would allow us to infer its luminosity and rule out certain classes of variable X-ray sources.

An estimate of the mass of the black hole involved will provide strong support for classification of a candidate transient as a WTDE (providing the LC shows anticipated WTDE-like behaviour). Recently, a tidal disruption event module \citep{mockler2018weigh} has been developed for MOSFiT \citep{Guillochon2018mosfit}, an open source Python-based code for Bayesian parameter estimation of astronomical transients based on semi-analytical fitting of their multi-wavelength lightcurves. This extension module provides estimates of $M_{\mathrm{bh}}$ (amongst other parameters) involved in MS TDEs. However, a module based on MS TDEs may not be appropriate to use for WTDEs, since the uncertainty surrounding their X-ray emission (in particular, the extent of super-Eddington accretion these events go through at flaring) may significantly bias any parameter estimates constructed from inferred posterior distributions. A further MOSFiT module specialised to WTDEs may require development (potentially in light of eROSITA's discoveries). We note our ability to constrain $M_{\mathrm{bh}}$ from the X-ray lightcurve data will also be limited by eROSITA's time sampling; the most accurate parameter estimates will likely follow from candidates that we can track for consecutive days in eROSITA's polar fields, or for which follow-up X-ray coverage with other missions can be compiled.  

An alternate way to estimate $M_{\mathrm{bh}}$ takes advantage of the majority of WTDE hosts expected to be DGs. For DGs with bulge mass estimates available, we may also crudely estimate $M_{\mathrm{bh}}$ from extrapolation of the $M_{\mathrm{bh}}-M_{*}$ relation \citep{McConnell2013}. Since DGs are typically low luminosity (and hard to identify in surveys), and assuming that the majority of DGs are satellites of more massive galaxies, then WTDEs will be observed as off-galactic centre transients.  If a new, flaring X-ray source originating from a DG is detected, deeper follow-up observations with either \textit{Chandra} or \textit{XMM-Newton} would provide a higher quality X-ray spectrum than eROSITA and be very useful for constraining the X-ray emission processes for the event (for example, are there components to the X-ray spectrum other than a black-body accretion disc?). Such follow-up would also provide better sampling of the X-ray light curve, allowing us to study the time evolution of $\dot{M}_{\text{acc}}$ for these systems and better understand the physical processes involved in the WTDE itself.

MS TDEs will significantly outnumber WTDEs across the DG population. However, it should be possible to distinguish between these two TDE classes as their LCs are expected to differ significantly within the IMBH mass range. As a basic comparison of LC behaviour, we consider a TDE for two cases: i) a $1M_{\odot}$ WD with $R_{\text{wd}}=0.01R_{\odot}$, and ii) a $1M_{\odot}$ MS star with $R_{\text{ms}}=1R_{\odot}$. For a configuration where $10^5M_{\odot}$ BH with $\beta=1$, $t_{\text{fb}}$ scales $\propto R^{3/2}_{*}$ (equation~\ref{eqn:t_peak}) such that the decay timescale for the MS TDE is a factor of 1000 greater than for the WTDE (and similarly also for $t_{\text{peak}}$). The slowing of MS TDEs from BHs $<10^6M_{\odot}$ is also found in \citet{Guillochon2015}, whereby these events typically evolve over several years due to highly inefficient debris circularisation and large viscous timescales for these systems. This is in agreement with a recent potential MS TDE identification from an IMBH \citep{Lin2018}, which was found to decay over a decade. 
We also highlight here that eROSITA should identify a set of MS TDE candidates from IMBHs that will require subsequent X-ray follow-up observations to track LC decay after its all-sky survey finishes.  

Approximately one in every six WD TDEs could result in a thermonuclear transient \citep{Macleod2016}. Their optical lightcurves are anticipated to be 1a-like, but may differ by possessing shorter rise times (10-12 days), maximum luminosities $\sim 8 \times 10^{42}$erg s$^{-1}$ (dimmer than 1a) and single-peaked in the near-infrared bands, unlike standard 1as (eg. \citealt{Kasen2006SecondarySupernovae}). These events may already have been observed as the calcium-rich gap transients \citep{Sell2015}, though no significant X-ray counterpart has been found for these. eROSITA's All-Sky Survey should coincide with wide area sky scans such as the \textit{Zwicky Transient Facility} \citep{Bellm2014}; finding a reported 1a-like supernova as a counterpart to an observed X-ray flare will support WTDE classification and help determine an origin of these Ca transients. \citet{Macleod2016} predict optical spectra for these 1a-like events with expected signatures: Doppler shifts $\sim 10^4 \text{kms}^{-1}$, P-Cygni lines, intermediate mass element production and a high dependence on viewing angle. For WTDEs that do not trigger thermonuclear burning, we anticipate optical spectra (based on MS TDEs observations eg. \citealt{Gezari2012,Arcavi2014}) to be mainly black body with different broadened emission lines depending on the composition of the WD being disrupted. For example, for WDs below 0.5$M_{\odot}$ we would expect broadened He emission lines to dominate the optical spectrum. 

\subsection{Uncertainties in WTDE rate estimates}\label{sec:caveats}
\subsubsection{TDE modeling}
As discussed in Section~\ref{sec:tdes}, mass fallback rates at early times post-disruption are predicted to be highly super-Eddington. We modelled emission during this period as Eddington-limited, such that the luminosity of the event will be determined by $M_{\mathrm{bh}}$ and approximately constant until sub-Eddington accretion rates are achieved. Several other classes of variable/ transient sources have been observed to be undergoing super-Eddington accretion (eg. ultra-luminous X-ray sources, \citealt{Walton2013}; and AGN, \citealt{Middleton2011}). However, the main factors that allow accretion to be super-Eddington are currently unclear. 
\citet{Evans2015} model accretion of the debris of a main sequence star onto an IMBH post-disruption. An initial phase of hyper-accretion is expected from accretion of portions of the star that falls directly onto the BH, followed by an approximately constant accretion rate until the debris is fully consumed\footnote{An instance of this hyper-accretion behaviour could be in \citet{Jonker2013}, where the lightcurve of the candidate WTDE shows very similar behaviour to the accretion rate curves plotted in Fig.~2 of \citet{Evans2015}.}. If WTDEs are capable of attaining super-Eddington accretion rates, it is likely they will be observed with eROSITA as a set of fast decaying transients detectable to higher $z_{\text{max}}$ anticipated in this work (if eROSITA can catch them very close to peak flaring).

In addition both WTDEs and MS TDEs are predicted to launch relativistic jets, with the potential to reach much higher luminosities. Furthermore, their lightcurves may fully trace the fallback rate over time, as opposed to Eddington limited phases such as in Fig.~\ref{fig:multi_wd_lcs}. These should be detectable to much higher $z$ than their non-jetted counterparts and boost the detection rate of WTDEs. We anticipate jetted WTDEs may be distinguished from their non-jetted counterparts via their harder X-ray spectra from Comptonisation of the quasi-blackbody radiation from the jet's photosphere (although our ability to distinguish between these two spectral models may be limited in cases of low source count rates) \citep{Shcherbakov2013}.  

On the other hand, the WTDE detection rate may be lowered by reprocessing of X-rays by optically thick outflows, predicted to be launched for cases of super Eddington fallback rates. \citet{Dai2018} propose a unified model of TDEs, whereby the viewing angle of the TDE affects whether one observes optical or X-ray dominated emission due to the angular dependence of the outflow's density. When viewed face on, one probes X-ray emission from the inner accretion disc, whereas a larger amount of X-ray radiation is reprocessed (due to increasing photoelectric absorption in the outflow density) with increasing viewing angle. Whilst the authors study super-Eddington accretion in MS TDEs, it is reasonable to assume that a similar viewing-angle dependence may also be present for WTDEs.

For simplicity, we have neglected black hole spin in this work. Introducing spin modifies the innermost stable orbit (ISCO) around the BH; larger spins lead to smaller ISCO radii such that the tidal disruption radius lies outside the ISCO and tidal disruption flares are possible (as opposed to the star being swallowed whole). Thus for a given WD, the maximum $M_{\mathrm{bh}}$ that could tidally disrupt it is greater, allowing black holes up to $10^6M_{\odot}$ to disrupt WDs. This could represent an increase in the number of BHs with potential to fully disrupt WDs, depending on the net alignment of black hole spins in the local universe.

\subsubsection{Simulating eROSITA observations}\label{sec:sim_erosita_observations}
The simulated source populations in this work have only included WTDEs and AGN. Since unresolved AGN populations are expected to contribute most to the CXB (eg. \citet{Lehmer2012TheGalaxies}), realistically modeling the AGN component is necessary for accurate estimates of source detection efficiency. We do not expect exclusion of other X-ray source populations, such as stars and galaxy clusters, to significantly affect our detection efficiency estimates. Furthermore, galaxy clusters in the redshift range for which we are WTDE-sensitive should be clearly identified as extended sources. We also note that galaxy clusters may provide a good place to search for WTDEs due to their large assembly of stellar mass, including GCs and
DGs, in their dark matter potentials. 

In addition, we have not included any other sources of X-ray variability that will be observed with eROSITA. Whilst inclusion of variability might allow for a rough quantification of the false positive rate for WTDE detection, no robust models currently exist for simulating synthetic X-ray lightcurves of all classes of variable X-ray sources. Even if these were available, it would still be a highly non-trivial task to robustly quantify the false positive rate. Our approach allows us to isolate the issue of how frequently eROSITA will detect WTDEs, from the broader problem of how many WTDEs can be identified amongst eROSITA's variable source population (since eROSITA's sparse time sampling introduces further complication to variable source classification). We have reported estimated detection rates of WTDEs where detection is counted as the source being detected by eSASS, but note the number of identified WTDEs will differ from the estimated detection rate. 

\section{Conclusions}\label{sec:conclusions}
We have simulated an extensive set of eROSITA observations during its all-sky survey of WTDEs, incorporating a realistic, non-variable X-ray background consisting of a CXB, SXB and a particle background. eROSITA's detection sensitivity to WTDEs as a function of black hole mass, redshift and $\mathcal{T}$ was then inferred. We estimated a novel rate density for WTDEs from BHs with masses between $10^3$ and $10^5M_{\odot}$, which was then combined with the detection sensitivities to estimate the rate of eROSITA detecting WTDEs. By the end of its 4 year all-sky survey, eROSITA should be able to probe a sample of $\sim 3$ quiescent intermediate mass black holes involved in WTDEs, if the estimated luminosity function of low luminosity galaxies can be extended down to the lowest luminosity dwarf galaxies observed outside the Milky Way. Due to the higher WTDE rate densities anticipated for dwarf galaxies, we expect eROSITA to be more sensitive to detecting disruptions involving black holes with masses $>10^4M_{\odot}$. Most detected WTDEs will be found within a few days of flaring, and detectable up to $z<0.24$. 

In addition, we have explored and demonstrated the usefulness of the SIXTE simulator in forecasting transient detection abilities of future X-ray surveys; the Python scripts used in this work will be made available in the near future\footnote{\url{https://github.com/amalyali}}.  
\section*{Acknowledgements}
A.M. thanks Andrea Merloni for helpful feedback on an early draft of this work; Jeremy Sanders, Nicolas Clerc and the eSASS team (Hermann Brunner, Christoph Gro{\ss}berger) for sharing their Python scripts; Jacob Chitham and the SIXTE team (Thomas Dauser, J\"orn Wilms) for computational advice. A.M. acknowledges support from and participation in the International Max-Planck Research School (IMPRS) on Astrophysics at the Ludwig-Maximilians University of Munich (LMU). We also thank the anonymous reviewer for their useful comments.




\bibliographystyle{mnras}
\bibliography{detecting_wtdes_erosita.bib} 







\bsp	
\label{lastpage}
\end{document}